\documentclass[aps,prd,twocolumn,showpacs,nofootinbib]{revtex4}

\usepackage{graphicx}
\usepackage{dcolumn}
\usepackage{amsmath}
\usepackage{amssymb}

\begin{document}

\title{Cosmological Constraints on Bulk Neutrinos} 
\author{Kevork N. Abazajian$^{1,2}$}
\author{George M. Fuller$^1$} 
\author{Mitesh Patel$^1$} 
\affiliation{$^1$Department of Physics,
University of California, San Diego, La Jolla, California
92093-0319}
\affiliation{$^2$NASA/Fermilab Astrophysics Center, 
Fermi National Accelerator Laboratory, Batavia, IL 60510-0500}
\affiliation{\tt aba@fnal.gov, gfuller@ucsd.edu, mitesh@physics.ucsd.edu}
\date{May 1, 2002}

\begin{abstract}
Recent models invoking extra spacelike dimensions inhabited by (bulk)
neutrinos are shown to have significant cosmological effects if the
size of the largest extra dimension is $R \gtrsim 1 {\rm \,fm}$. We
consider effects on cosmic microwave background anisotropies, big bang
nucleosynthesis, deuterium and $^6$Li photoproduction, diffuse photon
backgrounds, and structure formation. The resulting constraints can be
stronger than either bulk graviton overproduction constraints or
laboratory constraints.
\end{abstract}

\pacs{98.80.Cq, 11.10.Kk, 95.35.+d, 14.60.Pq  \hspace{3cm}
  FERMILAB-Pub-02/075-A} 
 
\maketitle
In this Letter we describe several cosmological constraints on models
for neutrino mass which rely on bulk fermions propagating in compact
extra spacelike dimensions. Extra spacetime dimensions, long provided
under the \ae gis of Kaluza-Klein (KK) and superstring theories, have
played an essential role in recent attempts to solve fundamental
problems in particle physics~\cite{fundamentalphysics,arkani}. In
particular, in some theories invoking $n$ compact extra spacelike
dimensions, all Standard Model (SM) fields are localized on a
three-dimensional surface (3-brane), but gravity experiences the full
spacetime (bulk)~\cite{arkani}. This yields the relation $M_{\rm Pl}^2
= M_F^{n+2} V_n$ between the new fundamental $(4+n)$-dimensional
reduced Planck scale $M_F$ and the four-dimensional reduced Planck
scale $M_{\rm Pl} = (4\pi G_N)^{-1/2}$, where $V_n$ is the volume of
the additional dimensions. If the volume of the internal space is
sufficiently large, $M_F$ can be much smaller than $M_{\rm Pl}$,
giving rise to a low-scale theory of quantum gravity ({\it e.g.}, $V_2
\approx 5\times10^{-5} {\rm\,mm}^2$ gives $M_F\sim 10 {\rm\,TeV}$ for
$n=2$).

In this framework, if $M_F$ is sufficiently small, there is no longer
a heavy mass scale available in the theory to suppress neutrino masses
relative to other fermion masses via a seesaw or similar mechanism
\cite{seesaw}. Several higher-dimensional mechanisms have been
developed~\cite{dienesarkani}, however, which can give neutrino masses
and mixings consistent~\cite{models,barbieri,ramond} with the solar,
atmospheric, and accelerator neutrino experiments~\cite{nuexpts}. One
widely used scheme posits the existence of SM-singlet fermions
(neutrinos) which propagate in the bulk but couple via Yukawa
interactions with the SM-doublet (active) neutrinos on our brane. This
setup gives up to three light Dirac neutrino masses $\mu_i$ associated
with the active neutrino flavors $\nu_e$, $\nu_\mu$, and/or
$\nu_\tau$. In addition, each bulk neutrino appears on our brane as a
tower of massive KK modes ({\it i.e.}, sterile neutrinos), and the
vacuum mixing angle between an active neutrino and a mode with mass
$m_{\rm mode} \gg \mu_i$ is $\theta_{\rm mode} \simeq
\sqrt{2}\mu_i/m_{\rm mode}$. The mass distribution of the modes
depends on the geometry of the internal space. The simplest and most
widely adopted geometry of the internal dimensions is that of an
$n$-dimensional torus with radii $R_j$ ($1\leq j \leq n$), for which
the mode masses are $m^2_{\bf{k}} = k_1^2 / R_1^2 + \cdots + k_n^2 /
R_n^2$, where, in bulk neutrino models, ${\bf k} = (k_1,\cdots,k_n)$
is an $n$-tuple of whole numbers, and where we assume that the bare
masses of the bulk fermions are negligible.  Several authors have
found non-standard solutions to the neutrino anomalies in this
framework~\cite{models,barbieri,ramond}.  These solutions require
$R_1^{-1} \lesssim 1 {\rm \,eV}$ ($R_1 \gtrsim 0.1 \,\mu{\rm m}$) for
the largest dimension $R_1$, for otherwise they reduce to those for a
standard Dirac neutrino mass~\cite{nuexpts,solar}. We show how these
and non-toroidally compactified models with densely distributed KK
modes affect standard cosmology through their production in the early
universe and subsequent decay.

The incoherent production of sterile-KK neutrinos of mass $m_k$ in the
early universe is a nonthermal process governed by a Boltzmann
equation~\cite{dw,afp2}
\begin{equation}
\label{boltz}
\frac{d}{dt}f_k = \Gamma_{\alpha k} f_\alpha -
\frac{m_k}{E}\frac{1}{\tau_k} f_{k} + \sum_{l>k} {\rm C}_{k,l}[f_l],
\end{equation}
where $f_i=f_i(p,t)$ are momentum- and time-dependent distribution
functions, and where $\alpha$ is an active neutrino label and $k,l$
are mode labels of a specific KK tower. We discuss the case in which
$R = R_1$ is the radius of the largest extra dimension and all other
dimensions are small enough to have no effect on low energy neutrino
physics. We have ignored the flavor coupling of multiple towers.  The
first term in Eq. (\ref{boltz}) is the conversion rate from active to
sterile species and the second results from the decay of a mode with
lifetime $\tau_k$. The latter arises because singlet neutrinos which
mix with active neutrinos can decay either to SM or bulk states. On
the brane, the partial decay width of the process
$\nu_k\rightarrow3\nu$ is $\sin^2\theta_k G_F^2 m_k^5/192\pi^3 = G_F^2
m_k^3 \mu_i^2/96\pi^3$ and that of the radiative decay
$\nu_k\rightarrow\nu \gamma$ is smaller by a factor $27\alpha/8\pi$
\cite{barger}.  We have also included in our calculations the
contributions to $\tau_k$ from visible and hadronic decays estimated
from the partial decay widths of the $Z^0$ boson~\cite{pdg}.  In the
bulk, the $k^\prime$-summed width of the decay
$\nu_k\rightarrow\nu_{k^\prime} h_{k-k^\prime}$ is $\sim m_k^4 R /
12\pi M_{\rm Pl}^2$, where $h_{k-k^\prime}$ is a KK graviton mode
\cite{mohapatrainflation}. The last term in Eq. (\ref{boltz})
represents the decay contribution of all higher modes $l>k$ into mode
$k$, and ${\rm C}_{k,l}$ is the appropriate collision operator.

The conversion rate $\Gamma_{\alpha k}=\Gamma_{\alpha
k}(p,t)=(\Gamma/2)\langle P_{\alpha k}\rangle$ to KK modes is the
product of half the interaction rate $\Gamma$ of the neutrinos with
the plasma and the average probability $\langle P_{\alpha k}\rangle$
that an active neutrino $\nu_\alpha$ scatters into $\nu_k$.  The
probability depends on the matter mixing angle and the damping rate
$D=\Gamma/2$~\cite{stod}:
\begin{equation}
\label{avgprob}
\langle P_{\alpha k}\rangle\simeq{\frac{1}{2}}
\frac{\Delta_k^2\sin^22\theta_k}{\Delta_k^2 \sin^2
2\theta_k+D^2+(\Delta_k\cos 2\theta_k-V)^2}.
\end{equation}
Here, $\Delta_k \simeq m_k^2/2p$, and $V = V^L + V^T$ is the full weak
potential including lepton and thermal contributions.  We assume a
small lepton number of order the baryon number (so $V^L \ll V^T$),
since a larger lepton number serves only to enhance sterile neutrino
or anti-neutrino production.  Eq. (\ref{avgprob}) incorporates the
standard physically well-motivated two-neutrino active-sterile matter
mixing angle and the effects of quantum damping~\cite{stod}. Our
constraints depend on the deleterious effects of the relatively high
modes, in which regime this formalism is identical to that derived
from direct diagonalization of the tower mass matrix~\cite{barbieri}.
The finite temperature potentials $V^T$ from the neutrino and charged
lepton backgrounds of the same flavor are $(8\sqrt{2}G_F E_\nu/3
m_{\rm Z}^2)(\rho_\nu+\rho_{\bar\nu})$ and $(8\sqrt{2}\,G_F E_\nu/3
m_{\rm W}^2)(\rho_l +\rho_{\bar{l}})$, respectively~\cite{nr}.

Another important effect is the dilution of modes populated at
temperatures $T \gg 100 \rm\,MeV$. Disappearance of relativistic
degrees of freedom manifests as heating of the plasma relative to the
KK modes.  We include this effect by following separately the complete
time-temperature relations for the photons and modes.

We explore the cosmological ramifications of two representative
classes of bulk neutrino models: Class I where $M_\ast \gtrsim 250
{\rm\,TeV}$; and Class II wherein $M_\ast \sim 1-10 {\rm\,TeV}$.
Here, $M_\ast^{n+2} \equiv (2\pi)^n M_F^{n+2}$. In either Class the
active neutrinos may be coupled with one, two, or three bulk
neutrinos.  The cosmological overproduction of bulk graviton modes
limits temperatures in this scenario to be less than $T_\ast^g \simeq
10^{(6n-15)/(n+2)} {\rm\,MeV}(M_\ast/{\rm\,TeV})$ which is an upper
limit on the ``normalcy'' temperature $T_\ast$ at which the universe
must be free of bulk modes~\cite{arkphenom}.  Models falling into
Class I correspond to large $T_\ast^g$ ($\gtrsim T_{\rm electroweak}
\sim 100 {\rm\,GeV}$), while those in Class II have very low
$T_\ast^g$ ($\lesssim {\rm\,GeV}$). In many Class II models
effectively only the lightest active neutrino couples to one KK
tower. In either Class if the heaviest active neutrino couples to a
tower then $\mu_3 > \sqrt{\delta m^2_{\rm SK}} \approx 0.057
{\rm\,eV}$~\cite{superk}.  We calculate here the constraints that
arise in cosmologies that satisfy normalcy temperature requirements of
their respective class.  However, if the radiation dominated era was
never significantly above the decoupling temperature $T_{\nu{\rm dec}}
\sim 1{\rm\,MeV}$ of the active neutrinos in a very low reheating
scenario for inflation, then early universe constraints cannot be
placed on either bulk graviton or bulk neutrino production.

For each of these classes we have solved numerically Eq. (\ref{boltz})
with ${\rm C}_{k,l}=0$ for the population of the $N$ lowest modes,
with fully self-consistent temperature evolution of all relevant
species.  We have performed this calculation for a single KK tower;
additional towers can serve only to enhance cosmological limits.  The
height $N$ of the tower is the highest mode populated at the
appropriate $T_\ast^g$.  Since our calculations begin at the highest
temperature $T_\ast^g$ permitted by graviton overproduction limits,
any adverse cosmological effects we find will imply that the normalcy
temperature $T_\ast$ must be significantly lower in bulk neutrino
models than implied from graviton production alone.  We have
conservatively incorporated the effects of decays in the bulk by
assuming the decay products' mass-energy negligibly affects the
dynamics of the universe.  For a given momentum the number of modes
produced per active neutrino per log-interval of temperature
$\Gamma_{\alpha k}/H$ depends implicitly on the mode mass $m_k$ via
Eq. (\ref{avgprob}), where $H$ is the instantaneous Hubble expansion
rate~\cite{dw}.  However, as shown analytically in Ref.~\cite{dw}, for
a mode non-relativistic at the decoupling temperature $T_{\nu{\rm
dec}}$ of the active neutrinos, the energy density
(given by $m_k\Gamma_{\alpha k}/H$ integrated over $\ln T$ and the
active neutrino distribution) is {\it independent} of the mode number,
despite the dependence of the mixing angle on the mode mass.

This result assumes dilution is negligible and depends on the modes
not having decayed appreciably~\cite{dw,afp2}.  Under the latter
assumption, and with some simplifications, we can extend Eq.~(9) in
Ref.~\cite{dw} to obtain an analytic estimate $N_{\nu_k}({\rm BBN})
\sim 10^{-3}\ (\mu_i/1{\rm\, eV})^2 (g_\ast^{\rm f}/g_{\ast k}^{\rm
p})$ for the energy density at $T_{\nu{\rm dec}}$ in a single mode $k$
relative to that in an active neutrino species.  The ratio
$(g_\ast^{\rm f}/g_{\ast k}^{\rm p})$ approximates dilution effects.
The statistical weight in relativistic particles in the plasma at
$T_{\nu\rm dec}$ is $g_\ast^{\rm f}$, and is $g_{\ast k}^{\rm p}$ at
the epoch of maximal production of mode $k$.  (Roughly, this maximal
production epoch is related to mode mass as $T_{\rm max}\simeq
133\,{\rm MeV}(m_k/1\,{\rm keV})^{1/3}$~\cite{dw}.)  Our numerical
calculations follow in detail the simultaneous production, dilution,
and decay of all relevant modes of various energies, giving a
$N_{\nu_k}({\rm BBN})$ dependence on $k$ which is flat modulo the
effects of dilution and decay.

Population of KK modes in the early universe leads to a number of
unacceptable effects that provide for compelling constraints. Our
calculated cosmological constraints differ from those in
Ref.~\cite{barbieri}, but they complement the supernova limits of
Refs.~\cite{barbieri,ramond}, and the laboratory constraints
of Ref.~\cite{Davoudiasl:2002fq}.

\begin{figure}[t]
\includegraphics[width=3.4in]{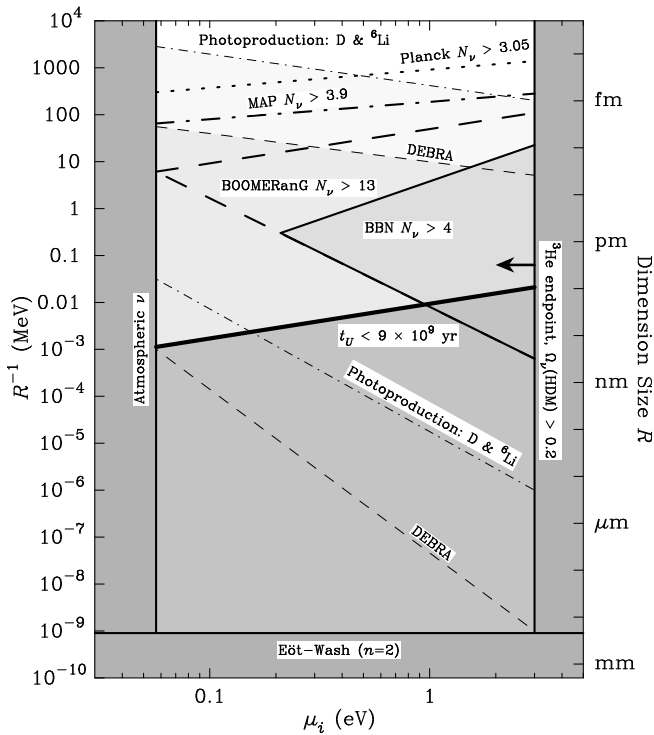}
\caption[]{
\label{fig1}
\small Cosmological constraints on Class I bulk
neutrinos. Photoproduction and DEBRA constrain regions between the
dot-dashed and short-dashed lines, respectively. The CMB constrains
parameters below the labeled BOOMERanG, MAP, and Planck lines and
above the long-dashed line. BBN constrains the region between the
light solid lines. Parameters must lie above the heavy solid line to
be consistent with the inferred age of the universe. The vertical
lines arise from the neutrino oscillation and $^3$H endpoint limits on
neutrino masses~\cite{nuexpts,solar,pdg,superk}.  Also
shown is the $218 {\rm\,\mu m}$ E\"ot-Wash limit on the size of two
congruent large extra dimensions~\cite{adelberger}.}
\end{figure}

Class I model constraints are given in Fig.~\ref{fig1}.  The total
effective number $N_\nu({\rm BBN})$ of neutrino flavors at the BBN
epoch must be less than that of 4 active neutrino species, since
otherwise the predicted and observed abundances of the light elements
are discordant~\cite{bbn}. We require the KK tower contribution $\sum
N_{\nu_k}({\rm BBN})$ to be less than that of a single active neutrino
flavor.
Photoproduction of deuterium (D) and $^6$Li due to decay of modes
after big bang nucleosynthesis (BBN)~\cite{photoprod} gives another
constraint.  Energetic cascades dissociate $^4$He into excessive
amounts of D~\cite{burtyt}.  The increase in energy density in
relativistic particles due to mode decay prior to cosmic microwave
background (CMB) decoupling can lead to suppression of the second CMB
acoustic peak.  The current limit is that the effective number of
neutrino flavors at decoupling is $N_\nu ({\rm CMB}) < 13$ at 95\%
certainty~\cite{hannestad}.  Measurements to higher multipole moments
by the Microwave Anisotropy Probe (MAP) (reaching $N_\nu ({\rm CMB})
\simeq 3.9$) and Planck (reaching $N_\nu ({\rm CMB}) \simeq 3.05$)
surveys will be able to further limit the relativistic energy present
at decoupling~\cite{lopezcmb}, or perhaps flag the fossil relativistic
energy of bulk modes at $R\sim 0.1 {\rm\,fm}$.  The increase in energy
density due to mode decays was found by summing the energy injected
between the neutrino and photon decoupling epochs. Another significant
constraint comes from the current limits on diffuse extra-galactic
background radiation (DEBRA) due to radiative decays of sterile
neutrinos occurring between CMB decoupling and today.  The photon
background so produced must have a total flux per unit solid angle
$d{\cal F}/d\Omega \lesssim (1{\rm\,MeV}/E)\rm\ cm^{-2}sr^{-1}s^{-1}$
\cite{dicuskolbteplitz,ressellturner,kolbturner}.  The expansion age
of the universe, $t_U > 9 \times 10^9\,{\rm yr}$, also provides a
constraint on the energy density in KK modes.  We have found that the
constraints from distortion of the CMB spectrum and the signals in the
solar neutrino experiments from mode decays are weaker than the
constraints above~\cite{dicuskolbteplitz,kolbturner,solar}.  Note that
the arguments above do not depend on the detailed mode structure but
rather on the existence of a high density of modes.

\begin{figure}[t]
\includegraphics[width=3.4in]{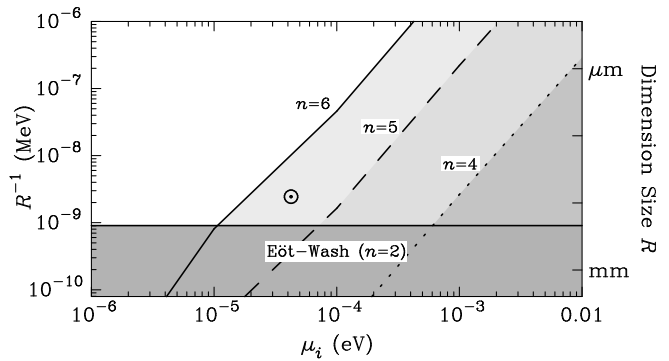}
\caption[]{
\label{fig2}
\small Structure formation constraints on Class II bulk neutrino
models with numbers of extra dimensions $n=6,5,4$ ($\Omega^{\rm HDM} >
0.1$ to the right of the slanted lines). By assumption, the size of
the largest extra dimension is $R$, and all other dimensions have sizes
$\ll {\rm pm}$.  For example, the solar neutrino solutions of Dvali \&
Smirnov and Caldwell, Mohapatra, \& Yellin~\cite{models} lie near
$\odot$.}
\end{figure}

Class II model effects are shown in Fig.~\ref{fig2}. Though high-lying
modes in these models are absent owing to low $T_\ast^g$, there may
remain enough energy density in low mass modes to comprise an
appreciable hot dark matter component. Structure formation
considerations~\cite{hdm} suggest that a hot component cannot
contribute $\Omega^{\rm HDM}> 0.1$. Contours of $\Omega^{\rm
HDM}_{\nu_s} = 0.1$ are shown in Fig.~\ref{fig2} for these models with
$n=6,5,4$ extra dimensions. Note that some recent models for solar
neutrino oscillations fall in a parameter range which could give an
appreciable $\Omega^{\rm HDM}_{\nu_s}$. Whether this can constitute a
true constraint depends on the precise relation between $T_\ast^g$ and
$M_\ast$ in these models~\cite{hallsmithfairbairn}. At present all
Class I and II models~\cite{models,barbieri,ramond} can escape
elimination by invoking a sufficiently low ($\lesssim 20 {\rm\,MeV}$)
re-heating temperature $T_r$ for inflation.

The spectrum of low-lying modes in Class II models could give a viable
dark matter candidate if $T_r$ is low.  Low $T_r$ results in
suppressed production of high-mass modes that provide the closure,
BBN, and decay constraints.  For low enough $\mu_i$, all modes
produced below $T_r$ survive until today and escape decay
constraints.  The possibility of a realistic dark matter component
from the KK modes is finely tuned. For instance, for $R^{-1} =
4\times10^{-8}\rm\,MeV$, and $\mu_i = 10^{-5}\rm\,eV$, $\Omega_{\nu_s}
\sim 0.1$ for $T_r \sim 1\rm\,GeV$, but $\Omega_{\nu_s} \sim 0.2$
for $T_r \sim 1.3\rm\,GeV$.  Albeit finely tuned, the latter case,
for which $\Omega^{\rm HDM}_{\nu_s} < 0.1$, is an interesting dark
matter candidate, comprising a mixture of hot, warm ($\sim {\rm
keV}$), and cold ($\sim {\rm MeV}$) components.

Modifications to Class I and II models may allow circumvention of our
constraints. A stronger dependence of the mixing angle on
$\mu_i/m_{\rm mode}$ would ensure that the population of the modes
would fall with increasing mass. An alternate dependence of mode
lifetime on $m_{\rm mode}$ and $\mu_i$ could eliminate some or all of
the constraints.  There could exist multiple additional (possibly fat)
branes in the bulk, devoid of energy density and parallel to our own,
onto which modes decay preferentially~\cite{arkphenom}.  If the
re-heating temperature $T_r$ of inflation is near $T_{\nu{\rm dec}}$,
no KK modes will be populated in the radiation dominated era, and
therefore the constraints presented here do not apply.  Some
population of KK modes can occur during reheating, or through resonant
production if there is a large lepton number, but we do not explore
these possibilities here. Also, the internal dimensions need not be
toroidally compactified~\cite{chm}.  A space which has a KK mode
decomposition with a sufficiently low mode density (or a {\it gap})
could evade cosmological constraints.  Bad effects of higher $\mu_i$
can be removed by restricting the KK towers to those built on low
$\mu_i$ ($\lesssim 10^{-4} {\rm\,eV}$) active neutrinos --- this is
what is (or should be) done in Class II models to avoid
constraint. However, models which make use of all three towers are in
some sense the most ``natural,'' albeit the most severely constrained.

Ultimately, our cosmological considerations may help to narrow the
otherwise prodigious range of parameters discussed by modelers to
date.

We acknowledge fruitful discussions with D.O.~Caldwell, E.~Gawiser,
K.~Intriligator, and especially R.N.~Mohapatra, and we thank the
Institute for Nuclear Theory, University of Washington, for
hospitality. This research was supported in part by NSF Grant
PHY00-99499 at UCSD, the DOE and NASA grant NAG 5-10842 at Fermilab,
and NASA GSRPs for KA and MP.


\begin{thebibliography}{}

\bibitem{fundamentalphysics} I.~Antoniadis, Phys. Lett. {\bf
B246},~377~(1990); J.D.~Lykken, Phys. Rev. D {\bf 54}, 3693 (1996);
E.~Witten, hep-ph/0002297, and references therein.

\bibitem{arkani} N.~Arkani-Hamed, S.~Dimopoulos, and G.~Dvali,
Phys. Lett. {\bf B429}, 263 (1998). I.~Antoniadis, N.~Arkani-Hamed,
S.~Dimopoulos, and G.~Dvali, {\it ibid.} {\bf B436}, 257 (1998).

\bibitem{seesaw} M.~Gell-Mann, P.~Ramond, and R.~Slansky, in {\it
Supergravity}, edited by P.~van~Nieuwenhuizen and D.Z.~Freedman
(North-Holland, Amsterdam, 1979); T.~Yanagida, in {\it Proceedings of
the Workshop on Unified Theories and Baryon Number in the Universe},
edited by A.~Sawada and A.~Sugamoto, K.E.K. preprint 79-18 (1979);
R.~N.~Mohapatra and G.~Senjanovic, Phys. Rev. Lett. {\bf 44}, 912
(1980).

\bibitem{dienesarkani} N.~Arkani-Hamed, S.~Dimopoulos, G.~Dvali, and
J.~March-Russell, Phys.\ Rev.\ D {\bf 65}, 024032 (2002); K.R.~Dienes,
E.~Dudas, and T.~Gherghetta, Nucl. Phys. {\bf B557}, 25 (1999);
K.R.~Dienes and I.~Sarcevic, Phys.\ Lett.\ B {\bf 500}, 133 (2001).

\bibitem{models} G.~Dvali and A.Yu.~Smirnov, Nucl. Phys. {\bf B563},
63 (1999); R.N.~Mohapatra, S.~Nandi, and A.~P\'{e}rez-Lorenzana,
Phys. Lett. {\bf B466}, 115 (1999); R.N.~Mohapatra and
A.~P\'{e}rez-Lorenzana, Nucl. Phys. {\bf B576}, 466 (2000);
R.N.~Mohapatra and A.~P\'{e}rez-Lorenzana, Nucl.\ Phys.\ B {\bf 593},
451 (2001); N.~Cosme {\it et al.}, Phys.\ Rev.\ D {\bf 63}, 113018
(2001); D.O.~Caldwell, R.N.~Mohapatra, and S.~J.~Yellin, Phys.\ Rev.\
Lett.\ {\bf 87}, 041601 (2001); {\it ibid.}, Phys.\ Rev.\ D {\bf 64},
073001 (2001).

\bibitem{barbieri} R.~Barbieri, P.~Creminelli, and A.~Strumia,
Nucl. Phys. {\bf B585}, 28 (2000).

\bibitem{ramond} 
A.~Lukas, P.~Ramond, A.~Romanino and G.G.~Ross,
Phys.\ Lett.\ B {\bf 495}, 136 (2000).

\bibitem{nuexpts} For a review, see, {\it e.g.}, S.M.~Bilenky,
C.~Giunti, and W.~Grimus, Prog. Part. Nucl. Phys. {\bf 43}, 1 (1999).

\bibitem{solar} 
J.N.~Bahcall, P.~Krastev,G.L.~Fogli, E.~Lisi, D.~Montanino and A.~Palazzo,
Phys.\ Rev.\ D {\bf 64}, 093007 (2001); 
J.N.~Bahcall, M.C.~Gonzalez-Garcia and C.~Pena-Garay,
JHEP {\bf 0108}, 014 (2001).

\bibitem{dw} S.~Dodelson and L.M.~Widrow, Phys. Rev. Lett. {\bf 72}, 17
(1994).

\bibitem{afp2} K.~Abazajian, G.M.~Fuller, and M.~Patel,
Phys.\ Rev.\ D {\bf 64}, 023501 (2001).

\bibitem{barger} V.~Barger, R.J.N.~Phillips, and S.~Sarkar,
Phys. Lett. {\bf B352}, 365 (1995); {\bf B356}, 617(E)
(1995).

\bibitem{pdg} D.E. Groom {\it et al.}, Eur. Phys. J.  {\bf C15}, 1
(2000).

\bibitem{mohapatrainflation} R.N.~Mohapatra, A.~P\'{e}rez-Lorenzana,
and C.A.~de S. Pires, Phys. Rev. D {\bf 62}, 105030 (2000).

\bibitem{stod} L.~Stodolsky, Phys.\ Rev.\ D {\bf 36}, 2273 (1987).

\bibitem{nr} D.~N\"otzold and G.~Raffelt, Nucl. Phys. {\bf B307}, 924
(1988).

\bibitem{arkphenom} N.~Arkani-Hamed, S.~Dimopoulos, and
G.~Dvali, Phys. Rev. D {\bf 59}, 086004 (1999).

\bibitem{superk} T.~Kajita  [Super-Kamiokande Collaboration],
Nucl.\ Phys.\ Proc.\ Suppl.\  {\bf 100} (2001) 107.

\bibitem{Davoudiasl:2002fq}
H.~Davoudiasl, P.~Langacker and M.~Perelstein,
Phys.\ Rev.\ D {\bf 65}, 105015 (2002).

\bibitem{adelberger} C.D.~Hoyle {\it et al.}, Phys. Rev. Lett. {\bf
86}, 1418 (2001).

\bibitem{bbn} R.H.~Cyburt, B.D.~Fields and K.A.~Olive,
Astropart.\ Phys.\  {\bf 17}, 87 (2002).

\bibitem{photoprod} See K.~Jedamzik, Phys. Rev. Lett. {\bf 84}, 3248
(2000), and references therein.

\bibitem{burtyt} S.~Burles and D.~Tytler, Astrophys. J. {\bf 499}, 699
(1998); {\bf 507}, 732 (1998).

\bibitem{hannestad} S.~Hannestad, Phys. Rev. Lett. {\bf 85}, 4203
(2000).

\bibitem{lopezcmb} R.E.~Lopez, S.~Dodelson, A.~Heckler, and
M.S.~Turner, Phys. Rev. Lett. {\bf 82}, 3952 (1999).

\bibitem{dicuskolbteplitz} D.A.~Dicus, E.W.~Kolb, and V.L.~Teplitz,
Astrophys. J. {\bf 221}, 327 (1978).

\bibitem{ressellturner} M.T.~Ressell and M.S.~Turner, Comments 
Astrophys. {\bf 14}, 323 (1989).

\bibitem{kolbturner} E.W.~Kolb and M.S.~Turner, {\it The Early
Universe} (Addison-Wesley, Reading, Massachusetts, 1993).

\bibitem{hdm} O.~Elgaroy {\it et al.}, Phys.\ Rev.\ Lett.\ {\bf 89},
061301 (2002); X.~m.~Wang, M.~Tegmark and M.~Zaldarriaga, Phys.\ Rev.\
D {\bf 65}, 123001 (2002); J.R.~Primack and M.A.K.~Gross,
astro-ph/0007165; E.~Gawiser, astro-ph/0005475.

\bibitem{hallsmithfairbairn} L.J.~Hall and D.~Smith, Phys. Rev. D {\bf
60}, 085008 (1999); M.~Fairbairn, Phys.\ Lett.\ B {\bf 508}, 335 (2001).

\bibitem{chm} N.~Kaloper, J.~March-Russell, G.D.~Starkman, and
M.~Trodden, Phys. Rev. Lett. {\bf 85}, 928 (2000).



\end{thebibliography}
\end{document}